\documentclass[11pt]{article}
\usepackage{graphicx}
\usepackage{amssymb}
\usepackage{amsmath}
\usepackage[parfill]{parskip}  

\title{Approaches for modeling magnetic nanoparticle dynamics}

\author{Daniel B Reeves [1] and John B Weaver [1,2]}

\date{\today}

\begin{document}

\maketitle

{\small [1] Dept. of Physics and Astronomy, Dartmouth College, Hanover NH 03755 USA}\\
{\small [2] Dept. of Radiology, Geisel School of Medicine, Hanover NH 03755 USA}

\begin{abstract}
Magnetic nanoparticles are useful biological probes as well as therapeutic agents.  There have been several approaches used to model nanoparticle magnetization dynamics for both Brownian as well as N\'{e}el rotation. The magnetizations are often of interest and can be compared with experimental results. Here we summarize these approaches including the Stoner-Wohlfarth approach, and stochastic approaches including thermal fluctuations. Non-equilibrium related temperature effects can be described by a distribution function approach (Fokker-Planck equation) or a stochastic differential equation (Langevin equation). Approximate models in several regimes can be derived from these general approaches to simplify implementation.\end{abstract}

\section{Introduction and nanoparticle applications}

Magnetic nanoparticles (MNPs) are useful in many biophysical and medical applications because they can be remotely controlled and monitored with magnetic fields. For these applications, it is important to model the dynamics of the particles; and many approaches have been used previously to do so. In this case `modeling' means simulating the magnetizations of the particles over time to study how different variables (e.g., field strength, anisotropy) affect their dynamics. In this review, we outline the major advances in MNP modeling including state-of-the-art techniques requiring numerical simulations and accounting for many experimentally verified phenomena. We also include approximate theories that admit analytical solutions, formulations that are easier to implement and other expressions that are more transparent.

\subsection{Sensing and imaging}

Magnetic nanoparticles can be used as sensors to detect properties of their local micro-environments when rotations couple MNP dynamics to the environmental parameters\cite{KOH,HAUN,CHUNG}. For example, the concentration of specific molecules can be detected because their dynamics can be noticeably different when bound and unbound\cite{XJ}. This is depicted in Fig.~\ref{apps}(a). Nanoparticle sensing is sensitive (down to 100pM of analyte and nanogram amounts of iron) and could potentially be used \emph{in vivo} to detect changing concentrations over time. Aside from molecular sensing, MNPs have been used as probes to measure local temperatures\cite{TEMPEST}, viscosities\cite{VISC}, and local environment rigidity\cite{RIGID}. MNPs have been used as MRI contrast agents\cite{SUNMRI}, and the developing technology of magnetic particle imaging (MPI). MPI uses the particles themselves as high-contrast imaging agents which hold the possibility for time-domain tracers\cite{MPINAT,ANNAC,MPI3D}.

\subsection{MNP hyperthermia}

\begin{figure}[h!]\begin{centering}\includegraphics[width=3.25in]{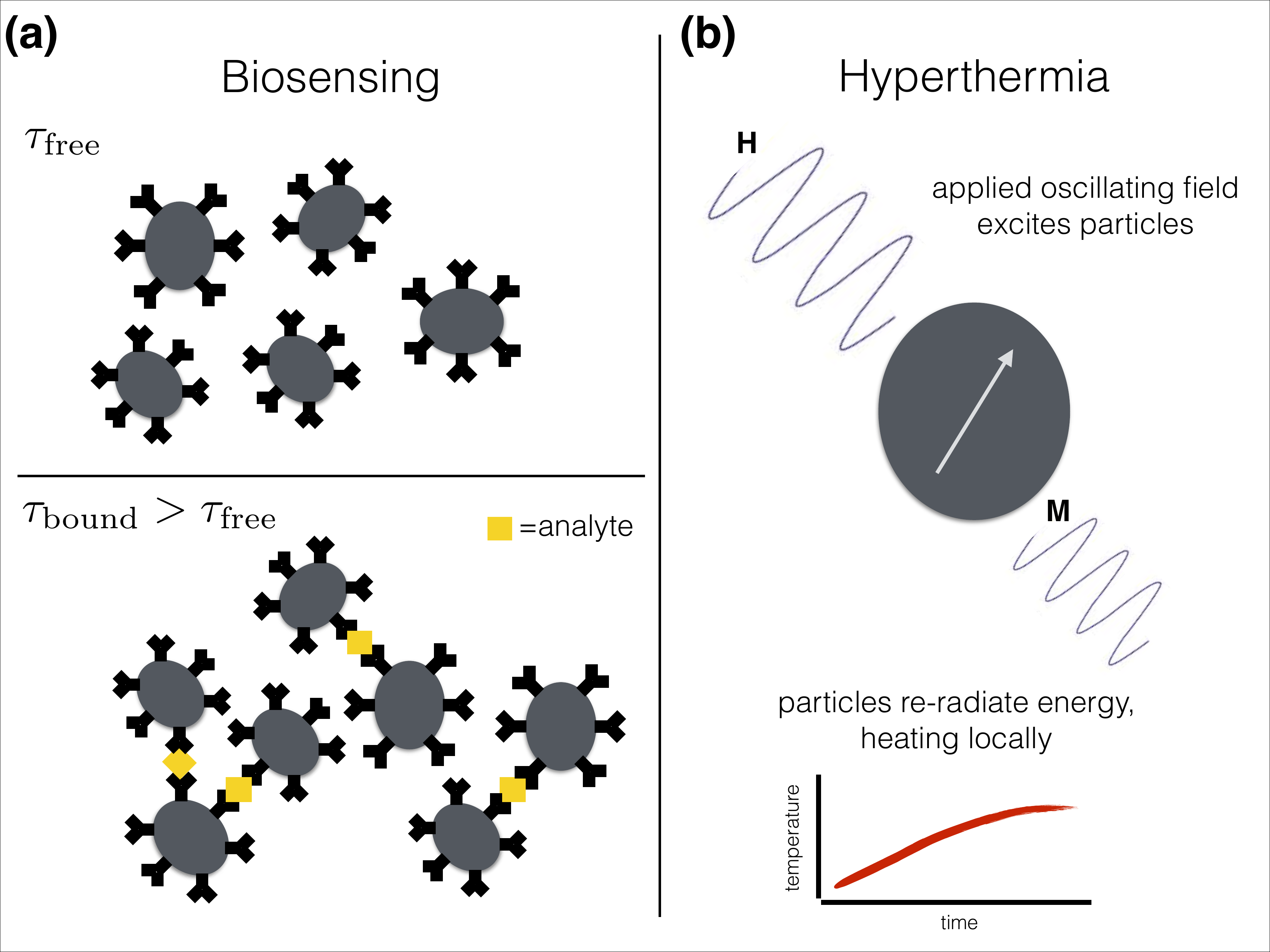} \caption{Two applications for magnetic nanoparticles (a) sensing specific `analyte' biomolecules, and (b) locally heating through dissipative losses from oscillating particles.} \label{apps}  \end{centering}\end{figure}

Magnetic nanoparticle (MNP) hyperthermia is considered a potentially useful addition to current cancer treatment modalities\cite{HERGT,JORDAN}. Nanoparticles can be directly injected\cite{GIUSTINI} or targeted biologically\cite{OMID} to achieve specific localized therapy. Furthermore, it appears cells ingest the MNPs, so cellular distribution may be possible\cite{RSC}. If oscillating magnetic fields are applied to the sample, diverse physical mechanisms prevent the particles from following the field exactly, and thus energy is dissipated locally. This energy can be thought of as heat, and it has been shown that cytotoxic heating is possible\cite{KMK,RINALDI}. The principle of MNP hyperthermia can be conceptualized as in Fig.~\ref{apps}(b). It is important to design efficient heating agents and much work has been done to characterize MNPs and improve properties such as magnetic moment magnitude and anisotropy energy\cite{GOYA,KMK}. 

\section{Nanoparticle rotations and relaxation times}

\begin{figure}[h!]\begin{centering}\includegraphics[width=3.25in]{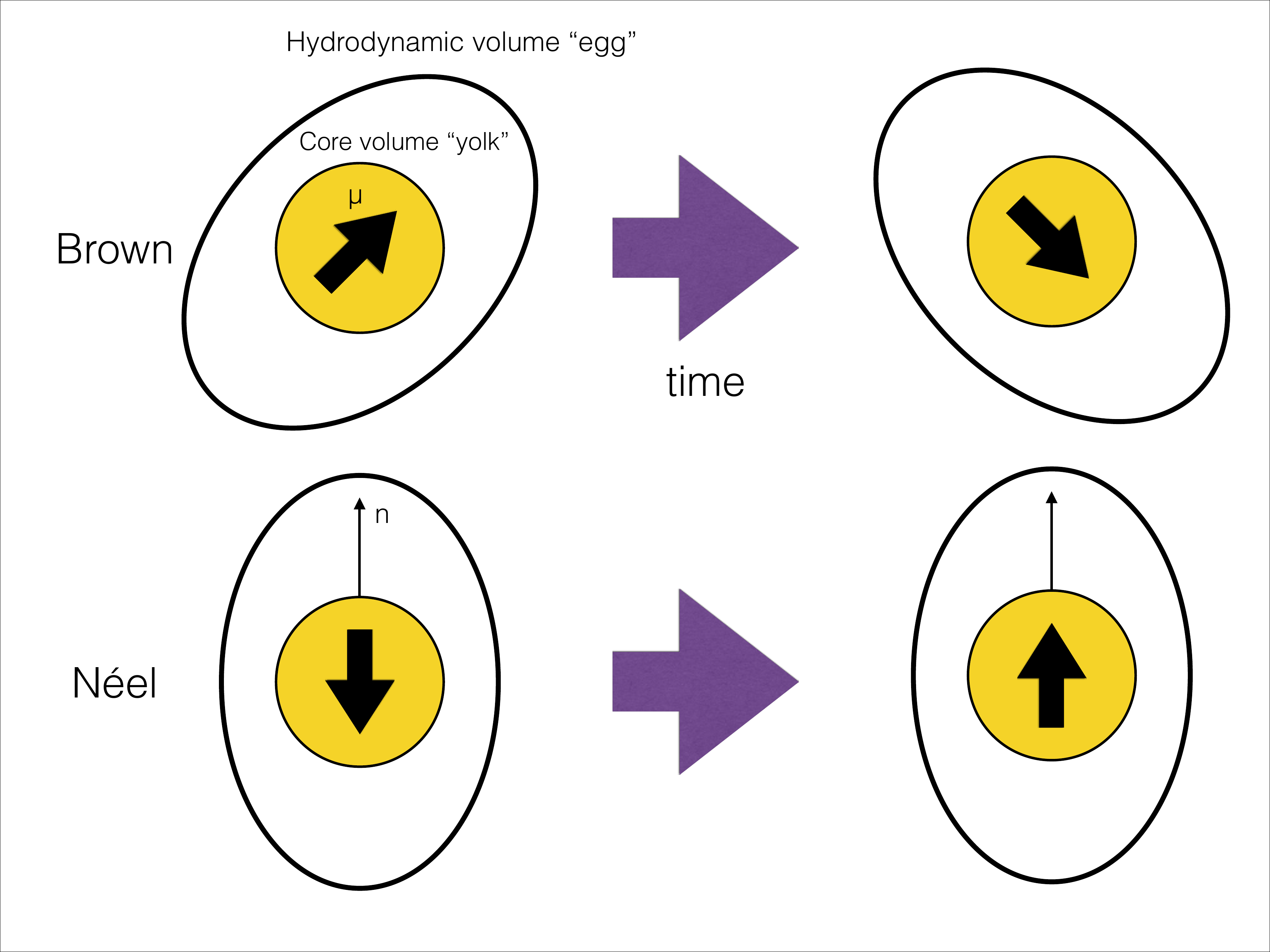} \caption{ N\'{e}el and Brownian particles relax differently. The moment can rotate internally to align with the anisotropy axis $n$, or the entire particle can rotate.} \label{eggyolk}  \end{centering}\end{figure}

There are two mechanisms for nanoparticles to rotate their magnetic moment $\mu$. This is visualized in Fig.~\ref{eggyolk}. The whole particle can rotate, as in so-called Brownian rotation, named for Robert Brown. The moment can rotate internally due to restructuring of electronic states. This so-called N\'{e}el rotation is named for Louis N\'{e}el, who first described the phenomenon while studying magnetic remanence in geological samples\cite{NEEL}. The timescale for a perturbed system to return to equilibrium is called the `relaxation time'. This expression is different for each mechanism and depends on the parameters of interest. 

\subsection{Brownian relaxation time}
The Brownian relaxation time, sometimes called the Einstein relaxation time\cite{EIN} based on Einstein's original work to derive the timescale from Brownian motion arguments is written
\begin{equation}\tau_B=\frac{3\eta V_h}{kT} \label{browntime}\end{equation}
in terms of the suspension viscosity $\eta$, the hydrodynamic volume of the particle $V_h$ and the thermal energy introduced with Boltzmann's constant $k$ and local temperature $T$.

\subsection{N\'{e}el relaxation time}
The N\'{e}el relaxation time derives from the thermal movements between two potential energy minima that arise from postulating a single anisotropy axis $n$ (see Fig.~\ref{eggyolk}). The anisotropy energy is written in terms of the anisotropy constant $K$ and the magnetic core volume $V_c$; the unitless ratio of this energy to the thermal energy is defined $\sigma$. Thus, the relaxation time can be expressed
\begin{equation}\tau_N=\tau_0e^{KV_c/kT}=\tau_0e^\sigma \label{neeltime}\end{equation}
where the `event time' $\tau_0$ is often quoted as $10^{-10}$s but is defined
\begin{equation} \tau_{0}=\frac{\mu}{2\gamma kT}\frac{(1+\alpha^2)}{\alpha} \end{equation}
in terms of the electron gyromagnetic ratio $\gamma=1.76\cdot10^{11}$Hz/T, a dimensionless magnetic damping parameter $\alpha$, and the magnetic moment---calculated using $M_s$ the material dependent saturation magnetization as $\mu=M_sV_c$. More precise forms of the N\'{e}el relaxation time including additional factors of $\sigma$ can be found in Ref.~\cite{FANNIN}. 

\subsection{WF relaxation time}
W F Brown developed a more general characteristic time including applied fields\cite{WFB}. Letting $\sigma\gg1$ he approximated this time to be\begin{equation}\tau_{\mathrm{WF}}=\frac{\tau_\mathrm{N}}{1-\epsilon^2}\frac{e^{\sigma\epsilon^2}}{\cosh\xi+\epsilon\sinh\xi} \end{equation}
where the applied magnetic field $H$ is accounted for with the unitless $\xi=\mu H/kT$. The ratio of anisotropic to magnetic energy is defined $\epsilon=\xi/2\sigma$ and the expression reduces to the equilibrium N\'{e}el time in the absence of an applied field.

\subsection{Relaxation time considerations}

An assumption found frequently in the literature is that the more prevalent relaxation mechanism is the one with the shorter relaxation time. The process is then approximated as parallel so the effective relaxation time is 
\begin{equation}\frac{1}{\tau_\mathrm{eff}}=\frac{1}{\tau_N}+\frac{1}{\tau_B}. \end{equation}
To the best of our knowledge there has been no theoretical work that deals with N\'{e}el and Brownian relaxation simultaneously, so currently this expression is not derived from first principle and does not incorporate the fact that the processes could be coupled \cite{MAMI,IVKOV}. Another consideration is that these relaxation times only hold for equilibrium conditions, and should not be applied without awareness of this constraint.

\section{Anisotropic hysteresis loops}

The energy of a magnetic particle with magnetization direction $\mathbf{m}$ and uniaxial  anisotropy direction $\mathbf{n}$ immersed in a magnetic field $\mathbf{H}$ can be written
\begin{equation} E=-KV_c\left(\mathbf{n}\cdot \mathbf{m}\right)^2-\mu\mathbf{m} \cdot \mathbf{H} \end{equation}
so that if the field is applied in the $\hat{z}$ direction, we define the angle from the polar to be $\theta$ and the angle between the moment and the easy axis to be $\phi$. Thus
\begin{equation} E(\theta,\phi)=KV_c\sin^2(\theta-\phi)-\mu H\cos\theta. \end{equation}
To find the minimum energy, the equilibrium state of the particle, with respect to the moment direction $\theta$ we write
\begin{equation} 0=\frac{\partial E(\theta,\phi)}{\partial \theta}=\mu H\sin\theta -2KV_c\sin(\theta-\phi)\cos(\theta-\phi) \end{equation}
or
\begin{equation} \epsilon=\frac{\sin\left[2(\theta-\phi)\right]}{2\sin\theta}. \end{equation}
To ensure a minima we also require $\frac{\partial^2 E(\theta,\phi)}{\partial \theta^2}>0$. Solving for the magnetization $M=\cos\theta$ and plotting this with respect to the normalized field $\epsilon$ we find that the so-called Stoner-Wohlfarth hysteresis loop emerges\cite{STOWO}.

This model is useful because the area of the loop indicates the energy dissipated when the particle is forced to change its orientation from one potential minima to another. This can then be used as a model for hyperthermic heating given particles with specific sizes, saturation magnetizations, and anisotropy constants. The weakness of the model is that it assumes no relaxational timescale for the process, thus neglecting that any phenomena depend on the frequency of the applied field. Still, models based on the theory have been extended to model heating more realistically\cite{CARREY}.

\section{Differential equations for the magnetizations}

There exist more general methods to model the behaviors of the magnetic moments in various applied fields.  This can be accomplished by describing the time dynamics of the magnetization using differential equations. Then the varying rotation methods as well as the specific conditions the particles experience in various applications can be simulated.

\subsection{Brownian dynamics}

A Brownian particle can be modeled phenomenologically with a balance of torques\cite{DBRSIM}:
\begin{equation}  \mathcal{T}=\underbrace{\mu \mathbf{m} \times \mathbf{H}}_\mathrm{magnetic}-\underbrace{6\eta V_{h} \mathbf{m} \times \frac{\partial\mathbf{m}}{\partial t}}_\mathrm{viscous}\label{Teff}.\end{equation}
We have neglected the acceleration term because the nanoparticles have a very small Reynold's number. Balancing this equation admits a differential equation for the magnetization which can be rewritten in terms of a unitless field $\boldsymbol{\xi}=\mu \mathbf{H}/kT$ and the Brownian relaxation time $\tau_B$ as
\begin{equation} \frac{d\mathbf{m}}{d t}= \frac{\left( \mathbf{m} \times \boldsymbol{\xi}\right) \times \mathbf{m}}{2\tau_B}. \label{dmm}\end{equation} 
This magnetization equation can be enhanced by including thermal fluctuations of the magnetization of the particles. There is no one general way to do this, and we will see the prescription for N\'{e}el particles is slightly different. Here we supplement the torques with the stochastic term $\mathcal{T}_{s}=\boldsymbol{\lambda}_t/\sqrt{\tau_B}$. The magnitude of this term depends on the relaxation time so that larger torques occur at higher temperatures and lower viscosities. We have also introduced the  white noise process $\boldsymbol{\lambda}_t$ which we define to have
\begin{equation} \langle \boldsymbol{\lambda}^i_t\rangle=0, \hspace{2mm} \langle\boldsymbol{\lambda}^i_t \boldsymbol{\lambda}^j_{t'} \rangle=\delta_{ij}\delta(t-t') \end{equation} 
where the process is uncorrelated in each cartesian direction $i,j\in x,y,z$ and is called white because its variance is a Dirac delta function in the time domain and thus is a constant in the frequency domain \cite{CHANDRA}. In a physical system, noise is not truly white but is a fair assumption when the stochastic torques have a much shorter timescale than the rotations of the particles. We write the stochastic differential equation, called a Langevin equation\cite{GARPAL} as
\begin{equation} \frac{d\mathbf{m}}{d t}= \frac{\left( \mathbf{m} \times \boldsymbol{\xi}\right) \times \mathbf{m}}{2\tau_B}+ \frac{\boldsymbol{\lambda}_t\times \mathbf{m}}{\sqrt{\tau_B}}. \label{dMeq}\end{equation} 
This equation is used to model Brownian particles and is amenable to changing local variables like fluid viscosity and temperature as well as externally applied fields.

 \subsection{N\'{e}el dynamics}
The equation for the change in the magnetization of a N\'{e}el particle is the phenomenological Landau-Lifshitz-Gilbert (LLG) equation. This can be derived from the Larmor precession of a spin in a magnetic field with an added velocity-dependent damping term \cite{GILB}. The normalized internal moment then rotates with
\begin{equation} \frac{\mathrm{d}\mathbf{m}}{\mathrm{d}t}= \frac{\gamma}{1+\alpha^2}\left[\mathbf{H} \times \mathbf{m}+\alpha\mathbf{m} \times \left(\mathbf{H} \times\mathbf{m}\right) \right]. \label{llg}\end{equation}

The effective field $\mathbf{H}$ is used to include additional dynamics beyond the applied field\cite{HAASE}. For example we include an externally applied field with amplitude $H_o$ and frequency $\omega$, a field from the anisotropy axis $\mathbf{n}$\cite{MPIANI}, and a stochastic field $\mathbf{h}$, so that
\begin{equation} \mathbf{H}=\underbrace{H_o\hat{z}\cos{\omega t}}_\mathrm{applied}+\underbrace{\frac{2KV_c}{\mu}(\mathbf{m}\cdot\mathbf{n})\mathbf{n}}_\mathrm{anisotropy}+\underbrace{\mathbf{h}(t)}_\mathrm{stochastic}. \label{heff}\end{equation}
In this case, the fluctuations are added to the model with an additional stochastic field, a white noise field with zero-mean and standard deviation parameterized by the nanoparticle variables with $\mathbf{h}(t)=\sqrt{\frac{2kT\alpha}{\mu\gamma}}\boldsymbol{\lambda}_t$.
Some work has been attempted to describe the effects of dipole fields, which can also be added to the effective field\cite{HAASE} or treated as a mean field\cite{FELDY}. But because it seems the effects are actually of detriment to the effectiveness of hyperthermia, we do not discuss them here.

\subsection{Numerical integration of Langevin equations}

With the addition of the stochastic terms, the differential equations cannot be integrated as a normal Riemann integral\cite{OKS}. In general we see that the dimensions of the stochastic terms are actually the square-root of time, so in fact it is more precise to instead write these equations as integral equations of the form
\begin{equation}\mathrm{d}\mathbf{m}=\int a \mathrm{d}t+\int b \mathrm{d}W_t\end{equation}
where the $W_t$ represents the Wiener process, the integral of white noise and also called Brownian motion---the continuous analogue to the random walk. Additional information on the various types of stochastic differential equations can be found in Ref.~\cite{OKS,PLAT}. The usual method is to integrate the differential equation to first or second order, using the Euler-Marayuma or the Heun integration scheme, respectively\cite{GARD,DBRSIM}. Then, magnetization moments are found from successive solving \begin{equation}\langle \mathbf{m}^j \rangle = \frac{1}{N}\sum_i^N {\bf m}_i^j \end{equation} so that the average magnetization is when $j=1$.

\section{Distribution function approach leading to approximate models}

We saw in the Langevin equation formulation that it is possible to include thermal effects to models of nanoparticles. This is particularly useful when the particles are expected to be at room temperature (as in biological and medical applications). The finite temperature will cause magnetization fluctuations and thus the nanoparticles are accurately described as a distribution of states. Instead of repeatedly solving a stochastic differential equation, we can use a distribution function approach where the distribution of states evolves over time. 

\subsection{The Fokker-Planck equation}

We can represent a single nanoparticle's magnetization as a point on the unit sphere, so that the surface density of many magnetizations is determined by a function $f(\theta,\phi,t)$. As the particles rotate in space, the magnetization surface density changes, leading to a surface current $J$. The total number of nanoparticles is conserved, i.e., $\int f \mathrm{d}\Omega = 1$, defining the surface density function as a probability distribution. The normalization also implies the continuity equation
\begin{equation} \frac{\partial f}{\partial t} = -\nabla \cdot J. \end{equation}
The probability current depends on the probability itself; this is the key to understanding the dynamics. At temperature $T=0$ the current density $J$ depends only on the velocity ($\mathbf{v}=\frac{\mathrm{d}\mathbf{m}}{\mathrm{d}t}$) of the points on the sphere. However, when the temperature is not zero, the system evolves towards equilibrium. A postulated phenomenological `diffusion' of the distribution function\cite{WFB} accounts for the approach to equilibrium. This is represented by a new term in the continuity equation proportional to a diffusion constant $D$ and the gradient of the distribution function. We write the Fokker-Planck (FP) equation for the distribution function $f(\theta,\phi,t)$ as
\begin{equation} \frac{\partial f}{\partial t} = -\nabla \cdot \left[\frac{\mathrm{d}\mathbf{m}}{\mathrm{d}t} -D\nabla \right]f. \label{FP} \end{equation}
From this distribution, magnetization statistics can be determined using the definition of the probability moments 
\begin{equation} \int \mathbf{m}^j f(\theta,\phi,t) \mathrm{d}\Omega = \langle \mathbf{m}^j(t) \rangle.\end{equation}
What remains is to define the change in the magnetization and the diffusion constant. The magnetization dynamics are controlled by the zero temperature differential equations Eq.~\ref{dmm} and Eq.~\ref{llg}. The diffusion constant is determined by the parameters at equilibrium conditions, when there is no applied field and $\frac{\partial f}{\partial t} =0$. In general, solutions to the Fokker-Planck equation are not possible, so approximation methods are used in practice.

\subsection{N\'{e}el rotation and hyperthermia}

If the expression for the magnetization dynamics of N\'{e}el rotation is inserted, this FP equation can be used to describe the rotations of small particles that are fixed spatially. These are the expected conditions for nanoparticles during hyperthermia\cite{HAASE}. It is also possible to develop an analytical approximation based on the assumption that the distribution function is linear in the magnetization. This approximation has been used to model hyperthermia because the imaginary response indicates heat deposition\cite{ROSENWEIG}. Though these expressions are commonly used, we do not describe them here because recent work has shown that they are inadequate to fully model heating \cite{IVKOV,CARREY}. The methods are accurate but care is necessary to assure that they are applied in the correct range of validity: when magnetizations are indeed linear as when applied fields are weak, magnetic moments are small and/or frequencies are high\cite{VALID}.

\subsection{Brownian Fokker-Planck with cylindrically symmetric applied field}

Replacing the velocity of the magnetization in Eq.~\ref{FP} with Eq.~\ref{dMeq}, simplified slightly we have the FP for Brownian rotation
\begin{equation} \frac{\partial f}{\partial t} = -\nabla \cdot \left[\frac{\boldsymbol{\xi}-\mathbf{m} \left(\mathbf{m} \cdot \boldsymbol{\xi}\right)}{2\tau} - D\nabla \right]f.  \end{equation}
A general solution is not currently analytically possible, but if the applied magnetic field is in the $\hat{z}$-direction only (i.e. $\boldsymbol{\xi}=\xi\hat{z}$), the FP equation can be simplified. 

This involves writing out all the components and unit vectors of $\mathbf{m}$ in spherical coordinates, $$\mathbf{m}=\sin\theta\cos\phi\hat{x}+\sin\theta\sin\phi\hat{y}+\cos\theta\hat{z}$$ and each unit vector is expressed e.g., $$ \hat{x}=\nabla x=\left(\hat{\theta}\frac{\partial}{\partial\theta} +\hat{\phi}\frac{1}{\sin\theta}\frac{\partial}{\partial\phi}\right) \sin\theta\cos\phi$$ eventually we find $$\hat{z}-\mathbf{m}\cos\theta=-\sin\theta\hat{\theta}$$ where the second term completely cancels. Thus, we arrive at the intuitive result that the distribution function does not depend on the azimuthal angle $f=f(\theta,t)$ only. We now have the FP equation
\begin{equation} \frac{\partial f}{\partial t} = -\nabla \cdot \left[-\frac{\xi}{2\tau}\sin\theta f - D\frac{\partial f}{\partial \theta} \right]  \end{equation}
where by using the definition of the spherical gradient over only the polar angle we write
\begin{equation} \frac{\partial f}{\partial t} = -\frac{1}{\sin\theta}\frac{\partial}{\partial \theta}\left[\sin\theta \left(-\frac{\xi}{2\tau}\sin\theta f - D\frac{\partial f}{\partial \theta} \right)\right] \end{equation}
and lastly with the change of variables $x=\cos\theta$ we have $-\frac{1}{\sin\theta}\frac{\partial}{\partial \theta} = \frac{\partial}{\partial x}$ and with $D=1/2\tau$ we can write the 1-D FP equation,
\begin{equation} 2\tau\frac{\partial f}{\partial t} = \frac{\partial}{\partial x}\left[(1-x^2)\left(\frac{\partial f}{\partial x} - \xi f \right)\right]. \label{FP1D} \end{equation}
This equation has solutions in a Legendre polynomial expansion\cite{DEIS,FELDY}. But it is of interest to demonstrate a further approximation method, the so-called `effective field' method or `macroscopic relaxation equation'.

\subsection{Effective field model for low-frequency Brownian relaxation}

If we set the applied field to be a constant $\xi_0$ and let $\frac{\partial f}{\partial t} =0$, we find the normalized distribution function
\begin{equation} f(x) =  \frac{\xi_0}{4\pi\sinh\xi}e^{\xi x}.\end{equation}
This is the same result as we get from Boltzmann statistics (remembering $x=\cos\theta$ in terms of the polar angle because the field is applied in the $\hat{z}$ direction). We write the average magnetization in the direction of the field as $M$. Using this distribution function and integrating to get the first moment, we find, as we should, that the Langevin function $\mathcal{L}(x)$ describes the equilibrium dynamics of an ensemble of particles in an applied field:
\begin{equation} M = \mathcal{L}(\xi_0)=\coth\xi_0-1/\xi_0.\end{equation}
The valuable approximation is to assume the particles are always at equilibrium so that the distribution function retains the same functional form, but with a time varying field $\xi\rightarrow\xi(t)$. This `effective field' approach turns out to be very useful to model low frequency oscillating fields where the relaxation time of the particles is shorter than the period of the applied field\cite{FELDY,DBRSIM}. Multiplying by $x$ and integrating Eq.~\ref{FP1D} with the effective distribution and definitions of the probability moments leads to an equation for the second moment in terms of the first. We have then a differential equation for the average magnetizations at ``near-equilibrium'' (which is notoriously a slippery condition):
\begin{equation} \frac{\mathrm{d}M}{\mathrm{d}t} = -\frac{M}{\tau_B}\left(1-\frac{\xi(t)}{\xi_\mathrm{eff}}\right)\end{equation}
where $\xi_\mathrm{eff}$ is found from inverting the Langevin function at every step\cite{FELDY}. An inverse Langevin function can be used as in Ref.~\cite{PADE}. The effective field equation is useful for quicker analyses because it is much less computationally challenging than the full stochastic models.

\section{Summary}

We have seen that there are two mechanism for a magnetic particle to reorient its moment. The energy changes for particles to relax internally over energy barriers are also discussed in the framework of localized heating as is used in magnetic nanoparticle hyperthermia. The dynamics of these magnetic moments undergoing each relaxation mechanism separately are explored. The two most general approaches (the Langevin equation, or stochastic differential equation approach, and the Fokker-Planck, or distribution function approach) are illustrated. These include full variability of nanoparticle parameters as well as thermal fluctuations induced in a realistic setting. The methods for numerical integration are highlighted, and readers are directed to valuable sources. However, these equations are difficult and time consuming to solve, so we also introduce a macroscopic relaxation equation that is useful for low-frequency Brownian simulations.

\section{Acknowledgments}
The authors gratefully acknowledge the support of the Neukom Graduate Fellowship and NIH grant NIH-NCI 1U54CA151662-01 for funding.
	
\bibliographystyle{unsrt}
\bibliography{GRBE}

\end{document}